# Quantum Superlattices, Wannier Stark Ladders and the "Resonance" technique

C. D. Papageorgiou, A. C. Boucouvalas, and T. E. Raptis

*Abstract*— We present a new method for solving the Schrodinger equation using the Lossless Transmission Line Model (LTL). The LTL model although extensively used in fiber optics and optical fiber design, it has not yet found application in solid state problems. We develop the transformation theory mapping the wave equation to LTL and we apply the model to the case of a solid state periodic lattice. We extend the theory with an additional Wannier-Stark term and we show with results the flexibility and the strength of the technique. The advantages of the method for arbitrary potentials are also stressed.

I. INTRODUCTION

Periodic lattices are a very important topic in a wide range of applications and technologies, but there has been strikingly successful applications of such devices in photonic communication system components especially in lasers and optical fibre gratings to mention just a few. In recent papers work on the analysis of either periodic lattices of quantum wells [1] or artificial optical periodic lattices [2] with standard textbook methods [3] has been presented. Extending periodic lattices to the case of linear superlattices also has been studied in [4] under the presence of an external field causing the appearance of a Wannier-Stark ladder [5]. Here we review a previously introduced non-traditional method for the study of quantum wells and periodic lattices with or without the Wannier-Stark terms using for the first time the Transmission Line Resonance (TLR) technique [6], [7], [8].

Every type of quantum well, periodic or not as well as every optical planar structure are governed by second order partial differential equations of one variable and for every such equation there exists an exact correspondence with an equivalent lossless TL of which the eigenvalues and eigenfunctions can be calculated by a simple algorithm resulting in a very short computer code.

Furthermore, the shape of the quantum well potential function $V(x)$ can give valuable information about the region where the eigenvalues are confined. While the traditional analytic methods work well only for a limited set of potential functions e.g., square pulse shaped lattices or trigonometric functions the TLR can give accurate results in all possible cases. We believe that the approach presented here serves as a very good alternative for researchers in this area for cases where simplicity and speed of calculation is important.

In the next section II, we review some generic important properties of lossless transmission lines [9], [10]. In section III, we present the formulation of TLR for periodic lattices and superlattices [11], [12]. In the same section, the method is applied in the case of some types of periodic quantum wells as well as the case of an external static polarization field leading to the Wannier – Stark effect. Discussion and conclusions follow in section IV.

II. LOSSLESS TL AND THEIR PROPERTIES

Following standard electrical engineering textbooks, we represent an arbitrary non-homogeneous TL along a single dimension *x* via the pair of complex functions $\{Z(x), Y(x)\}$ also known as the "*Impedance*" and "*Admittance*" per meter. We also denote with $\{V(x), I(x)\}$ the "*Voltage*" and "*Current*" of the TL. The TL equations then obtain the form

$$\begin{cases} dV(x)/dx = -Z(x) I(x) \\ dI(x)/dx = -Y(x) V(x) \end{cases} \quad (1)$$

The ordinary differential equation (ODE) system in (1) is also equivalent to the 2nd order linear differential equation also known as the canonical Sturm – Louville form for the boundary value problem [13],[14]

This paragraph of the first footnote will contain the date on which you submitted your paper for review.

C. D. Papageorgiou is an associate professor of Electrical Engineering At the National Technical University of Athens, Greece (e-mail: chrpapa@central.ntua.gr).

A. C. Boucouvalas, is a professor of Communication Networks and Applications at the University of Peloponnese, Tripolis, Greece. He is also a visiting professor at Northumbria University, UK. (e-mail: acb@uop.gr).

T. E. Raptis is with the Computational Applications Group, at the Division of Applied Technologies of the National Center for Science and Research "Demokritos", Athens, Greece (e-mail: rtheo@dat.demokritos.gr).

$$\frac{d}{dx}\left[\frac{1}{Y(x)}\left(\frac{dI(x)}{dx}\right)\right] = Z(x)I(x) \quad (2)$$

The above is simply based on identifying the voltage function with $V(x) = -(1/Y(x))dI(x)/dx$ from the 2nd of (1) and replacing in the first of (1). In the TL case, boundary conditions will be given by the current-voltage values at terminal points $\pm L$ as $\{V(\pm L), I(\pm L)\}$.

A lossless TL is characterized by having the complex functions $\{Z(x), Y(x)\}$ being purely imaginary. We recall at this point, that in standard engineering practice, imaginary values generally represent the so called "*reactive*" part of the field which remains stored being normal to the propagation axis while the real part stands for the part of the field of which the energy gets dissipated in ohmic elements if present. The great interest of the lossless TL model stems from the fact that such models can always be tuned for specific values of a given parameter (usually frequency, energy, or wavenumber) which are the TL "eigenvalues". For any such eigenvalue, a specific pair of the functions $\{V(x), I(x)\}$ is formed which may then be identified with an "eigenfunction". In the present review we are interested in application of such tunable lossless TL models for which we may now write naturally the conditions $\{Z(x) = jX(x), Y(x) = jY(x)\}$ where $X(x)$ and $Y(x)$ are some real functions of $x$. Next, we concentrate on uniform infinitesimal parts of an inhomogeneous TL model extending in the interval $[x-dx/2, x+dx/2]$. For every such infinitesimal interval now we will have

$$\begin{cases} dV(x)/dx = -jX(x)I(x) \\ dI(x)/dx = -jY(x)V(x) \end{cases} \quad (3)$$

From the general analysis of TL equations we have the equivalent Transfer Matrix (TM) formalism for approximating the evolution of the Voltage and Current functions across successive infinitesimal intervals given as

$$\begin{bmatrix} V(x+dx/2) \\ I(x+dx/2) \end{bmatrix} = \mathbf{W} \begin{bmatrix} V(x-dx/2) \\ I(x-dx/2) \end{bmatrix}$$

$$\mathbf{W} = \begin{bmatrix} \cosh(\gamma dx) & Z\sinh(\gamma dx) \\ \frac{1}{Z}\sinh(\gamma dx) & \cosh(\gamma dx) \end{bmatrix} \quad (4)$$

In the above we identify the parameter $\gamma^2(x) = -X(x)Y(x)$ with the well-known "propagation function" of wave mechanics and the characteristic impedance with $Z(x) = \gamma(x)/jY(x) = \sqrt{X(x)/Y(x)}$. The definition above results in $\gamma$ function having two alternating branches of either real or purely imaginary values.

Considering that at a certain point the currents and voltages are known we can calculate them in every point of the TL. Each infinitesimal element can be interpreted as a small homogeneous TL which then become equivalent to a T-circuit as in figure 1. We again identify the local impedances as

$$\begin{cases} Z_B = Z\tanh(\gamma dx/2) \\ Z_P = \dfrac{Z}{\sinh(\gamma dx)} \end{cases}$$

Such infinitesimal circuits with $|\gamma(x)|dx \ll 1$ can be fairly well approximated with

$$Z_B = j\frac{\gamma^2 dx/2}{Y(x)} = -jX(x)dx/2$$

$$Z_P = \frac{j}{Y(x)dx}.$$

As the maximal value of $|\gamma|$ is by definition known a priori for all $x$ in an interval of interest, one can always define the x interval partition from a condition of the form $\max|\gamma(x)|\delta x < c, \ 0 < c \ll 1$. If need be, one can also make the same partition to be a function of $x$ in order to have a well defined value everywhere inside the interval.

The input reactance of any T-circuit $jX_{IN}$ is then immediately related to its output reactance $jX_{OUT}$ (that is equal to the input reactance of its next T-circuit) by just computing the total composite impedance using the simple circuit diagram of figure 1

$$jX_{IN} = Z_P(jX_{OUT}+Z_B)(jX_{OUT}+Z_B+Z_P)^{-1}+Z_B \quad (5)$$

Substituting the originally defined values for the various impedances in (5) leads to a simple recursive form as a partial fraction expansion

$$jX_{IN} = Z_B + \cfrac{1}{Z_P^{-1} + \cfrac{1}{Z_B + jX_{OUT}}}$$

We notice that repeated application of the above in our algorithm leads to a Continuous Fraction Expansion that has been recently associated with an interesting new type of geometry called the "*Chain of Horospheres*" [15].

With the above analysis we can now interpret the overall TL as a series of connected T-circuits. Bringing a TL into resonant state requires a tuning condition which is given from the demand that the total left and right reactances at any point are equal and opposite. An infinite TL is equivalent to its characteristic impedance. Thus, assuming reactances are known at the two terminal points $\pm L$, the reactances at any point $x$ in the interval $-L < x < L$ can be calculated left and right of $x$ for any $x$ inside this interval. Then the tuning condition for any given TL assuming appropriate frequencies or their associated wavenumbers for the parameter $\xi$ which becomes an eigenvalue at the roots, takes the form

$$f(\xi) = X_{left}(\xi) + X_{right}(\xi) \qquad (6)$$

For each eigenvalue $\xi$, the corresponding eigenfunction $y(x)=I(x)$ is then easily generated via the transfer matrix (4) starting from the terminal point $L$ with $I(L) = 1, V(L) = X(L)$.

In general, the same method is applicable in any case of optics, wave and quantum mechanics or any other physical phenomena ruled by either a 2$^{nd}$ order ODE or a separable PDE of which one degree of freedom (as for instance, the radial equation of a spherically symmetric Hydrogen atom) obtains the standard Sturm-Liouville form

$$\frac{d}{dx}\left[g(x)\left(\frac{dy}{dx}\right)\right] = f(x)y \qquad (7)$$

The generic equivalence with a TL model is then described by the equivalent ODE system

$$\begin{cases} dV(x)/dx = -jf(x)I(x) \\ dI(x)/dx = -\dfrac{j}{g(x)}V(x) \end{cases} \qquad (8)$$

with the identification

$$I(x) = y(x)$$
$$V(x) = jg(x)dI(x)/dx \quad .$$

Next we give some examples of application of the RTL method in generic quantum mechanical systems..

### III. APPLICATION IN THE CASE OF PERIODIC WELLS AND SUPERLATTICES IN THE SCHRÖDINGER PICTURE

The general treatment of a periodic structure, either of ordinary lattices or superlattices is given in terms of the Schrödinger equation. Following standard textbooks, we have the generic time independent equation for a particle of mass $m$ and $\hbar$ the reduced Planck's constant ($h/2\pi$)

$$\nabla^2 \Psi(\mathbf{r}) = [u(\mathbf{r}) - \varepsilon]\Psi(\mathbf{r})$$
$$u(\mathbf{r}) = \frac{2m}{\hbar^2}U(\mathbf{r}), \varepsilon = \frac{2mE}{\hbar^2} \qquad (9)$$

Restricting attention to the 1-D case we get the ODE

$$\frac{d^2 y(x)}{dx^2} = [u(x) - \varepsilon]y(x) \qquad (10)$$

We define $I(x) = y(x)$ and $V(x) = jdy(x)/dx$ so that (9) can be rewritten in the form of the ODE system for an LTL as

$$dV(x)/dx = -j(u(x) - \varepsilon)I(x)$$
$$dI(x)/dx = -jV(x) \qquad (11)$$

Defining again $\gamma^2 = U(x) - \varepsilon$ and Y=1 leads to the final form as

$$dV(x)/dx = -\frac{\gamma^2(x)}{j}I(x)$$
$$dI(x)/dx = -jV(x) \qquad (12)$$

We now turn attention to the problem of an arbitrary periodic potential. In this particular case, it has been proven on very general grounds the so called Bloch theorem according to which the wavefunction contains a periodic modulating function known as a Bloch wave. We assume a homogeneous TL with an infinitesimal thickness $\Delta x$ extending in the range $[x - \Delta x/2, x + \Delta x/2]$ with an equivalent factor $\gamma(x)$. This corresponds to an infinitesimal homogeneous T-circuit (shown in fig 1) with impedances $Z_B \approx j\gamma^2 \Delta x/2$ and $Z_P \approx j/\Delta x$. Successive infinitesimal circuits of the above type can be connected to approximate the continuous functions $V(x)$ and $I(x)$ due to their correspondence with $dy/dx, y(x)$ respectively. Following the same generic approach explained in section 2, we need to tune the resulting lossless TL model according to the matching condition between left and right layers separating any chosen point.

Given appropriate boundary conditions represented by terminal impedances at $Z(x = \pm L)$ the roots of the characteristic function $f$ of equation (6) are then the energy eigenvalues of both the TL and the original Schrödinger problem. For each such eigenvalue $\varepsilon$ the corresponding eigenfunctions can be easily computed with the aid of the generic Transfer Matrix method as provided by the general theory of the general LTL equation in the form

$$\begin{bmatrix} V(x-\Delta x/2) \\ I(x-\Delta x/2) \end{bmatrix} = \mathbf{W} \cdot \begin{bmatrix} V(x+\Delta x/2) \\ I(x+\Delta x/2) \end{bmatrix}$$
$$\mathbf{W} = \begin{bmatrix} \cosh(\gamma \Delta x) & -Z(x)\sinh(\gamma \Delta x) \\ -\dfrac{1}{Z(x)}\sinh(\gamma \Delta x) & \cosh(\gamma \Delta x) \end{bmatrix} \qquad (13)$$

In (13) the characteristic impedance is given as $Z(x) = -j\gamma(x)$. For an infinitesimal TL we approximate the above assuming $|\gamma(x)|\delta x << 1$ with $\cosh(\gamma(x)\Delta x) \approx 1$, and $\sinh(\gamma(x)\Delta x) \approx \gamma(x)\Delta x$. This simplifies (13) in the final form

$$\begin{bmatrix} V(x-\Delta x/2) \\ I(x-\Delta x/2) \end{bmatrix} = \mathbf{W} \cdot \begin{bmatrix} V(x+\Delta x/2) \\ I(x+\Delta x/2) \end{bmatrix}$$

$$\mathbf{W} = \begin{bmatrix} 1 & -j\gamma^2(x)\Delta x \\ j\Delta x & 1 \end{bmatrix} \quad (14)$$

Starting with $I(x_R)=1, V(x_R)=Z(x_R)$ from the right terminal impedance and assuming both $V$ and $I$ are defined for all successive points $n\Delta x$ as well as the wavefunction $y(x) = I(x)$ and its derivative.

We first apply the above in the simple case of a superlattice formed by a set of successive planar layers of alternating material along $x$ with constant quantum potentials $V_1 < V_2$ with alternating widths $d_1, d_2$ and total period $d_1 + d_2$ forming a so called superlattice with a periodic square well potential. We want to find the eigenvalues (eigenenergies) and eigenfunctions for a finite number of periods representing N similar dyadic blocks of material. The set must then be terminated at two barriers of height $V_2$ satisfying $Z(x_L) = Z(x_R) = j\sqrt{V_2 - \varepsilon}$. We also set $\gamma^2(x) = U(x) - \varepsilon$ and $Z=j\gamma(x)$. An example of a superlattice of ten such wells with $d_1 = 0.8, d_2 = 0.2, V_1 = 0, V_2 = 10$ together with two of their computed eigenfunctions is shown in figure 2. The $x$ interval was subdivided into 2000 infinitesimal subintervals with $\Delta x = 0.005$ and $|\gamma(x)| < \sqrt{10}$, thus their product $\max|\gamma(x)|\Delta x < 0.0167$ and for $p = |\gamma(x)|\Delta x$ $\sinh(p) \cong p$, $\tanh(p/2) \cong p/2$ and $\cosh(p) \cong 1$. Details of the algorithm used for this and the rest of the examples are provided in Appendix A.

Next we consider an example of a superlattice of a non squared potential with periodic wells with curvature and period $d=1$ as defined by

$$U(x) = \begin{cases} V_2[(1-\sin((x\pi/2)^2)^{-1} -1]/Ys, & 1-D/2 > x > D/2 \\ Ys = (1-\sin(D\pi/2)^2)^{-1} -1, & x < D/2 \text{ or } x > 1-D/2 \end{cases} \quad (15)$$

The superlattice is again considered to be terminated to the max. values $j\sqrt{V_2-\varepsilon}, V_2=10$. We also have that $\max|\gamma(x)|\Delta x < 0.0167$, so that for $p = |\gamma(x)|\Delta x$ the relations $\sinh(p) \cong p$, $\tanh(p/2) \cong p/2$ and $\cosh(p) \cong 1$, are valid.

In figure 3 the calculated eigenfunctions for the graded superlattice potential for two of the calculated eigenvalues are shown. The presented method is much simpler than existing alternatives leading to the same results. However, in the case of periodic wells with arbitrary curvature there is no general method for attacking this and similar problems. Also for a general periodic or even non periodic superlattice with an arbitrary graded potential, we know that there is no analytic expression of its eigenfunctions and thus no analytic approach for calculating eigenvalues.

As a last example of such a difficult case where standard treatment is usually given via the so called "perturbation method" [16], we apply RTL in the case of an externally applied electrical field across the x direction leading to the formation of the so called *"Wannier-Stark ladder"*. The Stark effect is well known in the literature and is responsible for shifting the energy levels due to polarization by a constant electric field applied between the two ends of a periodic lattice. In this case we assume a perturbation of the form $U'(x) = U(x) + E_0 x$ where $E_0$ is the intensity of the external DC electric field such that we now have

$$\gamma^2(x) = U'(x) - \varepsilon = U(x) + E_0 x - \varepsilon \quad (16)$$

In this case the original symmetry of the potential is broken causing the appearance of the characteristic ladder shapes. For each one of these ladders there is a possibility for the establishment of a set of eigenfunctions with their associated eigenvalues.

We can assume that for an adequate number of wells in the ladder, the terminal reactances at the starting and ending points are equal to the higher potential of the ladder or any higher value potential in order to confine the eigenfunctions inside the ladder. Following the previous analysis, i.e. dividing the Wannier-Stark ladder in a large number of very thin lattices $\Delta x$, the eigenvalues and their respective eigenfunctions can be calculated. In figures 4 and figures 5 two ladders are sown of squared and non squared wells together with two of their computed eigenvalues and their corresponding eigenfunctions for $E_0=1$. In this case we took the terminal impedances as $j\sqrt{V_2 + x - \varepsilon}, V_2 = 10, x = 10$ in this case to have for 2000 infinitesimal subintervals with $\Delta x = 0.005$ thus $\max|\gamma(x)|\Delta x < 0.0223$, so that for $p = |\gamma(x)|\Delta x$ the relations $\sinh(p) \cong p$, $\tanh(p/2) \cong p/2$ and $\cosh(p) \cong 1$, are valid again.

We can define a search region in the interval $[\varepsilon_1, \varepsilon_2]$ where one may take the limits to be the lower minimum and the upper maximum of any ladder well in order to find possible eigenfunctions related to this well. As we can notice the method of calculation for superlattices or Wannier-Stark ladders is exactly the same. The same resonance method can be applied also in any periodic or non periodic structures of any form of similar or dissimilar wells.

IV. DISCUSSION AND CONCLUSIONS

The above examination showed the ability of our approach to locate the eigenvalues from arbitrary potentials with ease. The significance of the particular type of Wannier-Stark effect extends beyond just the stationary states examined here. In another approach which is also non-perturbative [17] the problem is examined in the case of additional *ac* fields causing chaotic scattering. This situation is much worse in the case of abrupt high power and high voltage electrical impulses followed by strong current surges.

Recent theoretical and experimental work by the authors [19] [20] [21], showed the appearance of certain extreme phenomena even at the low energy limit that call for further study in terms of a full quantum mechanical treatment. Complete examination of such phenomena will be provided in a forthcoming publication.

We note in passing that while the problem of the transition from the microscopic to the macroscopic reality has been partially answered with the well known Ehrenfest theorem [18], there are still some obscured areas associated with the passage to macroscopic dissipative structures as for instance, in the case of determination of the current form in macroscopic radiating volumes or antennas from first principles. As an example we can take the case of a long thin antenna of length L under electrical excitation causing a polarization field in the interior of the bulk lattice.

We may consider the case of an initially uncharged thin, long metal wire or strip with a large conductivity. We can also simplify the problem by taking a free electron gas model where the influence of the periodic potential is negligible but the boundary conditions are to be considered important for taking into account correctly the finite length of the conductor. We are then confronted with a situation much like the case of a finite square well where the potential of the conducting electrons is practically very large in comparison with that of the underlying lattice wells. Hence, the whole wire can be seen as a kind of quantum trap for the electron gas.

In this practically infinite well only the known sinusoidal wavefunctions would be of importance with zeroes at positions $\pm L/2$. Ignoring at the moment the Stark effect associated with polarization, we first consider the statistical effect of a very large numbers of electrons of which every eigenfunction when added up becomes equivalent to the macroscopic harmonics for the overall electric current in the conductor with a frequency $\omega = 2L/c$.

Assuming an initial transient charging takes place that excites a number of harmonics in the wire which now acts as a temporary antenna. The externally applied electrical field can be interpreted as an overall Stark term superposed on the square well causing a strong polarization which subsequently causes an excess in the concentration of the electrons on the surface of the conductor.

This overpopulation of electrons is expected to have as a secondary effect the appearance of a new potential function of which the exact form is difficult to know yet we can approximate by a simple argument as follows. The huge transient internal potential in practice can also affect the various lattice atoms and this influence can be approximated as a wall of charge against a lattice atom with its internal electron shells due to the very large scale difference. In such a case a local atom will see a practically homogeneous field which will cause an additional temporary Stark effect breaking the symmetry of the lattice atomic potentials and raising the possibility of naked nuclei due to a retraction of the internal shell orbitals due to a very strong local Stark effect.

A hypothesis to be tested at a later stage regards the possibility of high power excitations of conductors to be able to cause even nuclear transmutations due to the baring of the lattice nuclei at least with some probability to be estimated in future studies. The already observed evidence from previous experimentation where complete disintegration and strong white light emission takes place suggests that the above is a possibility that must be tested more thoroughly.

The algorithm presented here is a very practical alternative to other more computationally heavy methods like *Hartree-Fock* and it is the opinion of the authors that can be extended even in such difficult cases as the transient excitation of long thin wires with a single or multiple materials as in superlattices. Existing codes as presented in App. A appear to converge rapidly and with a practically linearly increasing accuracy with respect to the coarse graining parameter δ$x$. Hence they may prove beneficial to the research community in a multiplicity of other similar cases.

APPENDIX

A set of MATLAB® codes that were used to produce the figure is provided together with a brief description of their use. In all applications shown in this paper a partition of the main interval in 2000 layers was used with a thickness $\Delta x$ =0.005 each. Thus taking into consideration that $Max(\gamma^2(x)) < 10$ for the superlattices $(\gamma^2(x))\Delta x < 0.0167$ and for the Wannier-Stark ladder for $E_0$=1 where $Max(\gamma^2(x)) < 20$ and $(\gamma^2(x))\Delta x < 0.0223$. Thus in both cases for $p = (\gamma^2(x))\Delta x$ the relations sinh(p)≅p, tanh(p/2)≅p/2 and cosh(p)≅1, are valid.

The eigenvalue finder is given by the bloch22 routine. The function $u(x)$ and the $E_0$ value are defined from any appropriate external interface function one of which for non squared wells is given by bloch11 rourine. MATLAB provides the facility of root finding functions for locating specific values as roots of the array returned by bloch22. The bloch33 routine is giving the eigenfunction of a potential function (superlattice or Wannier-Stark ladder) for a given eigenvalue. Resulting eigenvalues for all four cases examined in the main paper are provided in Table 1.

```matlab
function y=bloch11(x)
% Wannier Stark ladder for w3>0, v2=max potential of
superlattice for w3=0
% The wells have a width of 1
global  w3 v2
w2=.8;
w1=w3*x;

x1=x+1;
if x1>0;xx=x1-fix(x1);else xx=1+x1-fix(x1);end
xx=2*xx-1;

ys=(1/(1-(sin(w2*pi/2))^2)-1);
y=(1/(1-(sin(xx*pi/2))^2)-1)*v2/ys;

y=y+w1;
if xx>=w2;y=v2+w1;end
if xx<=-w2;y=v2+w1;end

function y=bloch22(e)
% root finder equation of Wannier Stark ladder terminated
% for the higher potential value of zz at the boundaries
% N1=number of wells of the ladder or superlattice w3=0

global N1 w3

N=2000;
dz=N1/N;
aa=bloch11(N1);
zz=j*sqrt(aa-e);
z1=zz;
for n=1:N;
    x=N*dz-(n-1)*dz-dz/2;
    cc=bloch11(x)-e;
    zb=j*cc*dz/2;
    zp=j/dz;
    zz=(zz+zb)*zp/(zz+zb+zp)+zb;
end
y=imag(zz+z1);

function y=bloch33(e)

% eigenfunction for Wannier Stark ladder of N1 wells for
given eigenvalue e
% terminated at barrier of its maximum potential

global  N1 tt t ttt
N=2000;
dz=N1/N;
as=sqrt(bloch11(N1)-e);
zz=[as;1];
f(1)=zz(2);
xx(1)=N1;

for n=1:N;
    x=N*dz-n*dz+dz/2;
    xx(n+1)=x-dz/2;
    cc=bloch11(x)-e;
    A=[1 cc*dz;dz 1];
    zz=A*zz;
    f(n+1)=zz(2);%eigenfunction
    ff(n+1)=bloch11(x);%superlattice or ladder

end

zz(1)/zz(2)
av=max(f);bv=min(f);
;%eigenfunction confined between 0 and e
f=(f-bv)/max(f-bv)*e
tt=f;t=xx;ttt=ff+.0001;
plot(xx,f,xx,ff+0.0001);
```

REFERENCES

[1] L. Hoddeson, E. Braun, J. Teichmann, S. Weart, *"Out of the Crystal Maze. Chapters from the History of Solid State Physics"*, Oxford Univ. Press, 1992.
[2] A. Smerzi, A. Trombetoni, "Optical Lattices: Theory," in *Emergent Nonlinear Phenomena in Bose-Einstein Condensates,* vol. 45 S p r i n g e r , 2008, pp. 247–265.
[3] C. Kittel, *Quantum Theory of Solids*. 2$^{nd}$ Ed., Willey, New York, 1987.
[4] C. Hamaguchi, *Basic Semiconductor Physics*, Springer, 2001.
[5] E. E. Mendez, G. R. Bastard "Wannier-Stark Ladders and Bloch Oscillations in Superlattices," *Physics Today*, vol. 46, no. 6, pp. 34–42, 1993.
[6] C. D. Papageorgiou, J. D. Kanellopoulos, "Equivalent Circuits in Fourier Space for the study of EM fields," *J. Phys. A: Math. Gen..*, vol. 15, pp. 2569–2580,  1982.
[7] C. D. Papageorgiou, A. D. Raptis, "A method for the solutions of Schrödinger equation," *Comp. Phys. Comm.*, Vol. 43, pp. 325 – 328, 1987.
[8] C. D. Papageorgiou et al., "A method for computing phase shifts for scattering,"  *J. Comp. App. Math.*, pp. 61 – 67, 1990
[9] L. N. Dvorsky, *Modern Transmission Line Theory and Applications*, R. E. Krieger Pub. Co.,  1979.
[10] G. Miano, A. Maffucci, *Transmission Lines and Lumped Circuits*, Sci. Direct, 2001.
[11] V. Grecchi *et al.*, "Stark Ladder of Resonances: Wannier Ladders and Perturbation Theory", *Comm. Math. Phys.* Vol. 159, No. 3, 1994, pp. 605-618.
[12] F. Meng, "Bloch oscillations and Wannier Stark Ladder study in Semiconductor Superlattice", PhD Thesis, HAL archives-ouvertes, Universite Paris-Sud – Paris XI, 2002, https://tel.archives-ouvertes.fr/tel-00795646
[13] H. Michiel, *Sturm-Liouville Theory*, Encyclopedia of Mathematics, Springer, 2001.
[14] G. Teschl, *Mathematical Methods in Quantum Mechanics with Applications in Schrödinger Operators*, Am. Math. Soc., 2009.
[15] A. F. Brandon, I. Short, "A Geometric Representation of Continued Fractions", *Am. Math.* Monthly, Vol. 121, No. 5,2014, pp. 391 - 402.
[16] A. D. Landau, E. M. Lifschitz, *Quantum Mechanics: Non-relativistic Theory*, 3$^{rd}$ Ed., Inst. Phys. Prob., USSR Academy of Sciences, Butterworth-Heinemann, 1977.
[17] M. Gluck *et al.*, "Wannier-Stark resonances in optical and semiconductor superlattices", *Phys. Rep.* Vol. 366, 2002, pp. 103-183.
[18] L. E. Ballentine, *Quantum Mechanics*, Prentice Hall, 1990
[19] C. D. Papageorgiou, T. E. Raptis, "Dipole Electromagnetic Forces on Thin Wires under Transient High Voltage Pulses," *Eur. Phys. J.: App. Phys.*, Vol. 48, 2009, pp. 31002-31006.

[20] C. D. Papageorgiou, T. E. Raptis, "A Solid State Ion Collider with Transient Current Pulses," Presented at CMMP 2010, IOP Conf. Series.

[21] C. D. Papageorgiou, T. E. Raptis, "Collisions in Metallic Lattices Under Transient Current Pulses," *Eur. Phys. J.: App. Phys.*, Vol. 54, No. 1, 2011, pp.

TABLE I
CALCULATED EIGENVALUES FOR TEN WELLS

| Sq. Wells | Sq. Wells/WS Ladder |
|---|---|
| 1.882323758928340 | 3.821083251932061 |
| 2.128029079655592 | 5.541936589278953 |
| 2.536350770548123 | 6.943427668535321 |
| 3.105290903104073 | 8.175341254729894 |
| 3.831524487427041 | 9.298748979739466 |
| 4.709576837991253 | 10.356808521701257 |
| 5.730039628915276 | 11.416307865400780 |
| 6.875187431560925 | 12.615851302654566 |
| 8.105853527282257 | 14.14932491397774 |
| 9.308110171804502 | 17.081006786119623 |
|  | 19.916402641206560 |

| Non Sq. Wells | Non Sq. Wells/ WS Ladder |
|---|---|
| 2.997490736501053 | 4.902143091788726 |
| 3.235745108757861 | 6.609827763390554 |
| 3.631388078391419 | 7.997876665249874 |
| 4.181977541483595 | 9.216355865107895 |
| 4.883435048859936 | 10.328372874665952 |
| 5.729013303641652 | 11.378556895144090 |
| 6.707050360468842 | 12.426419377119878 |
| 7.795465555034969 | 13.598121221788318 |
| 8.945606393470433 | 15.08727507151336 |
|  | 17.201288618849624 |
|  | 18.407750017160357 |

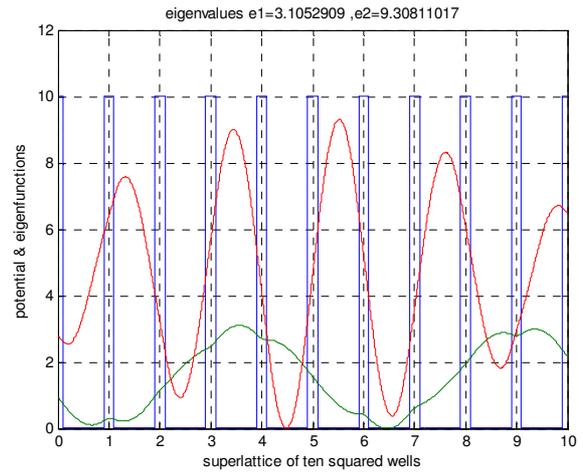

Fig. 2. Potential with corresponding wavefunctions obtained for ten periodic square wells via the Transfer Matrix for the 4th and 10th eigenvalues $e_1$ (green) and $e_2$ (red) respectively (see Table 1).

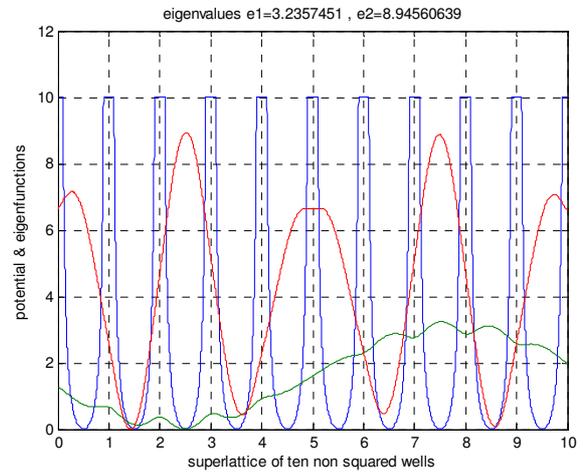

Fig. 3. Potential with corresponding wavefunctions obtained for ten periodic non square wells via the Transfer Matrix for the 2th and 9th eigenvalues $e_1$ (green) and $e_2$ (red) respectively (see Table 1).

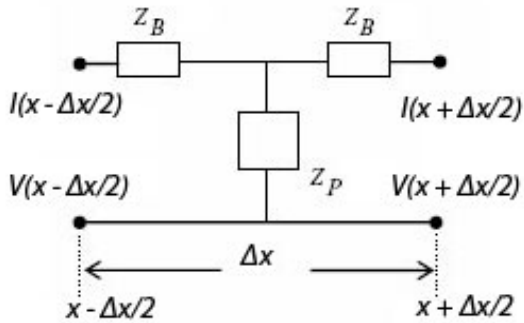

Fig. 1. The basic 2-ports T-circuit, representing a homogeneous differential element of a lossless transmission line.

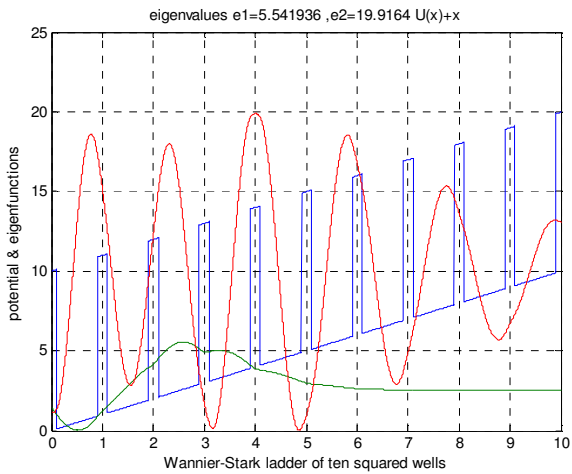

Fig. 4. Potential with corresponding wavefunctions obtained for Wannier-Stark ladder of ten square wells (Eo=1) via the Transfer Matrix for the 5th and 11th eigenvalues $e_1$ (green) and $e_2$ (red) respectively (see Table 1).

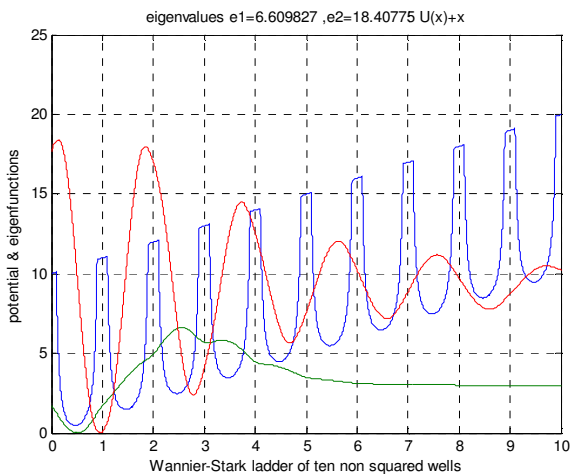

Fig. 5. Potential with corresponding wavefunctions obtained for Wannier-Stark ladder of ten non square wells (Eo=1) via the Transfer Matrix for the 2th and 11th eigenvalues $e_1$ (green) and $e_2$ (red) respectively (see Table 1).

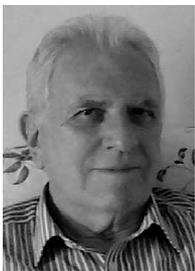

**Christos D. Papageorgiou** was born in Ioannina, Greece in 1943. He received the B.S. and M.S. degrees in electrical and mechanical engineering from the National Technical University of Athens, in 1966 and the Ph.D. degree in electrical engineering from Imperial College of London , in 1979. Assoc. Prof at the NTUA from 1979 and CEO in several public companies of which the most important were: Olympic Airways (1983-1985), & National Hellenic Railways (1987-1989 & 1996-1997). He was also top executive officer in a few private companies. Recently is the major shareholder of "Green Chimney Technologies P.C." that is operating in the area of Solar Energy.

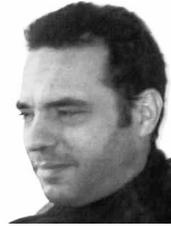

**T. E. Raptis** received the B.S. degree in physics from the University of Crete, Hreaklion, Greece in 1995. He is a Research Assistant with the Division of Applied Technologies, NCSR "Demokritos", Athens, Greece, as well as with the Power Electronics Sector of the Electrical Engineering Dept. in the Technological Educational Institute of Piraeus, Greece. His research interests include Physical Chemistry, Applied Electromagnetics, Complexity, Automata and Computational Structures, Machine Learning, Time Series Analysis, Dynamical Systems and Chaos.

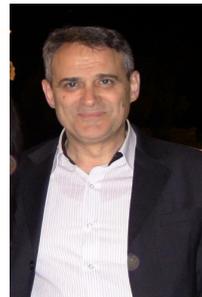

**Anthony C. Boucouvalas** (M'81-SM'00-F'02) is a Professor in Communications Networks and Applications at the University of Peloponnese in Tripoli, Greece. Prof. Boucouvalas has been actively involved with research in various aspects of fibre optic communications, wireless communications and multimedia and has an accumulated 35 years experience in well known academic and industrial research centers. He graduated with a B.Sc. in Electrical and Electronic Engineering from Newcastle upon Tyne University in 1978. He received his MSc and D.I.C. degrees in Communications Engineering, in 1979, from Imperial College, where he also received his PhD degree in Fibre Optics in 1982. Subsequently he joined GEC Hirst Research Centre, and became Group Leader and Divisional Chief Scientist until 1987, when he joined Hewlett Packard Laboratories as Project Manager. In 1994, when he joined Bournemouth University where he became a Professor in Multimedia Communications and Director of the Microelectronics and Multimedia research Centre at Bournemouth University. In 2007 he joined the Department of Telecommunication Science at the University of Peloponnese where he served for 6 years as Head of Department. His current research interests lie in optical wireless communications, fibre optic communications, inverse fibre optic problems, network protocols, and human-computer interfaces and Internet Applications. He has published over 310 papers. He is a Fellow of IET, (FIET), a Fellow of IEEE, (FIEEE), a Fellow of the Royal Society for the encouragement of Arts, Manufacturers and Commerce.